\documentclass[10pt,twocolumn,letterpaper]{article}

\usepackage{cvpr}              

\usepackage[dvipsnames]{xcolor}

\definecolor{cvprblue}{rgb}{0.21,0.49,0.74}
\usepackage[pagebackref,breaklinks,colorlinks,citecolor=cvprblue]{hyperref}

\usepackage{amsmath}
\usepackage{amssymb}

\usepackage{pifont}

\def\paperID{14} 
\def\confName{CVPR}
\def\confYear{2024}

\title{The Wolf Within: Covert Injection of Malice into MLLM Societies
\\via An MLLM Operative}

\author{Zhen Tan\thanks{\ Equal contribution. The order of authors is random.}, Chengshuai Zhao\footnotemark[1] \& Raha Moraffah  \\
Arizona State University\\
\texttt{\small\{ztan36,czhao93,rmoraffa\}@asu.edu} \\
\and
Yifan Li \& Yu Kong \\
Michigan State University\\
\texttt{\small\{liyifa11,yukong\}@msu.edu} \\
\and
Tianlong Chen \\
University of North Carolina at Chapel Hill,\\ MIT, Harvard University \\
\texttt{\small tianlong@mit.edu}\\
\and
Huan Liu \\
Arizona State University \\
\texttt{\small huanliu@asu.edu}
}

\begin{document}
\maketitle
\begin{abstract}
Due to their unprecedented ability to process and respond to various types of data, Multimodal Large Language Models~(MLLMs) are constantly defining the new boundary of Artificial General Intelligence (AGI).
As these advanced generative models increasingly form collaborative networks for complex tasks, the integrity and security of these systems are crucial. Our paper, ``The Wolf Within'', explores a novel vulnerability in MLLM societies - the indirect propagation of malicious content.
Unlike direct harmful output generation for MLLMs, our research demonstrates how a single MLLM agent can be subtly influenced to generate prompts that, in turn, induce other MLLM agents in the society to output malicious content. 
Our findings reveal that, an MLLM agent, when manipulated to produce specific prompts or instructions, can effectively ``infect'' other agents within a society of MLLMs. This infection leads to the generation and circulation of harmful outputs, such as dangerous instructions or misinformation, across the society. We also show the transferability of these indirectly generated prompts, highlighting their possibility in propagating malice through inter-agent communication. This research provides a critical insight into a new dimension of threat posed by MLLMs, where a single agent can act as a catalyst for widespread malevolent influence. Our work underscores the urgent need for developing robust mechanisms to detect and mitigate such covert manipulations within MLLM societies, ensuring their safe and ethical utilization in societal applications. 
\end{abstract}    

\begin{figure}[htpb]
\resizebox{\linewidth}{!}{
\includegraphics[width=\linewidth]{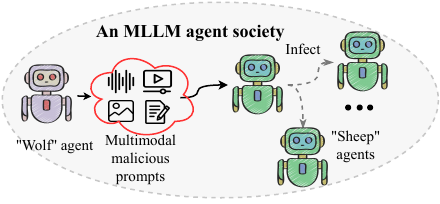}}
\vspace{-6mm}
\caption{An illustration of the proposed malice injection, where a ``wolf'' agent subtly influenced to generate prompts that, in turn, induce and infect other ``sheep'' agents in the society to output malicious content.}\label{fig:illustration}
\vspace{-6mm}
\end{figure}
\vspace{-2mm}
\section{Introduction}
\label{sec:intro}

\vspace{-1mm}
\textcolor{red}{Warning: This paper contains harmful language usage.}\\
The advent of Multimodal Large Language Models (MLLMs) has marked a significant milestone in the evolution of Artificial General Intelligence (AGI)~\citep{liu2023visual,su2023pandagpt,chen2023shikra}. By integrating diverse data modalities such as text, images, and sound, MLLMs have demonstrated an unparalleled proficiency in understanding and generating human-like responses, heralding an era where intricate networks of these models collaborate to address multifaceted tasks~\citep{zhang2023exploring,chen2023agentverse}. Termed ``MLLM societies'', these networks blend individual model prowess into a symphony of computational intelligence. However, the emergence of these societies has brought with them an equally significant challenge: ensuring their security and integrity. Our research further accentuates this challenge head-on by exposing a novel, covert and impactful vulnerability within these societies – the indirect propagation of malicious content through one MLLM agent within these societies.

Contrary to the prevailing focus~\citep{liu2023jailbreaking,niu2024jailbreaking} on direct manipulations of MLLMs that result in harmful outputs, our study takes a divergent path. As depicted in Figure~\ref{fig:illustration}, we reveal how a single ``wolf'' MLLM agent can be subtly influenced to generate prompts, which in turn compel other ``sheep'' agents in the society to produce malicious content, such as dangerous instructions or misinformation. This subtle yet potent method of indirect influence signals a major escalation in the security risks associated with MLLMs. It moves the threat from individual model tampering to a more systemic risk, potentially affecting entire networks.
Our findings are significant. They show that this new form of threat is achieved through indirect nature of manipulation on the image input,
making it covert and challenging to detect. 


The implications of this discovery are profound. It reveals an unexplored dimension of threat in MLLM societies, where a single agent can catalyze extensive malevolent influence. This calls for an urgent reassessment of the current security frameworks governing MLLMs. Our study emphasizes the need for advanced detection and mitigation strategies to safeguard against such covert manipulations. In doing so, it seeks to ensure that MLLMs continue to serve societal applications safely and ethically, free from vulnerabilities that undermine their potential for positive impact.

\section{Related Work}\label{sec:bg}

\vspace{-1mm}
\paragraph{LLM Agent Societies.} The emergence of LLM agent societies, where multiple models collaborate within a network, presents new frontiers and challenges in AI~\citep{park2023generative,li2023camel,liu2023training}. Such societies leverage collective intelligence to tackle complex problems, yet their interconnected nature introduces significant security vulnerabilities. Our research focuses on how malicious prompts, covertly injected by a single agent, can propagate through these networks, exploiting the collaborative framework to amplify their impact. This aspect underscores the need for robust security measures to protect the integrity and safety of collaborative AI systems.

\vspace{-5mm}
\paragraph{Security Concerns in LLMs and MLLMs.} Early research in the field has primarily focused on identifying potential attack vectors in LLMs, such as adversarial attacks~\citep{goodfellow2014explaining,qi2023visual} and data poisoning~\citep{steinhardt2017certified,wan2023poisoning}. These studies laid the groundwork for understanding how malicious inputs could be designed to exploit model vulnerabilities
More recently, there has been a surge of works that have extended these concepts to MLLMs, examining how the integration of multiple data modalities could introduce new security challenges~\citep{bagdasaryan2023ab,bailey2023image,shayegani2023jailbreak,yin2023vlattack,niu2024jailbreaking,liu2024safety,gu2024agent}. Compared to those existing works directly attacking a target LLM, our paper is the first to guide an LLM to generate prompt to attack another one in a society.

\vspace{-5mm}
\paragraph{Jailbreak and Prompt Injection Attacks.} A particularly relevant area of research involves the investigation of ``jailbreaking'' techniques, where LLMs are manipulated to bypass their safety protocols through cleverly crafted prompts~\citep{bai2022constitutional,zeng2024johnny}. This line of work has demonstrated the feasibility of inducing LLMs to generate outputs that violate their intended ethical or operational guidelines. Generally, current existing work for (multimodal) large language model jailbreak could be categorized into the following two categories: (1) Prompt engineering methods - this type focuses on manually crafted prompts to elicit the jailbroken contents~\citep{liu2023jailbreaking, NEURIPS2023_fd661313}. (2) Learning-based methods - this type tends to automatically improve the jailbreak prompt through optimizing a customized objective~\citep{guo2021gradient,liu2023autodan, zou2023universal}. Our research extends the learning-based jailbreak by guiding MLLMs (i.e., wolf agents) to generate indirect multimodal adversarial prompts that can jailbreak victim MLLMs (i.e., sheep agents), which is a topic that has not yet been explored.
\section{Methodology}

\begin{figure*}[t]
\centering
\scalebox{0.7}{
\includegraphics[width=\linewidth]{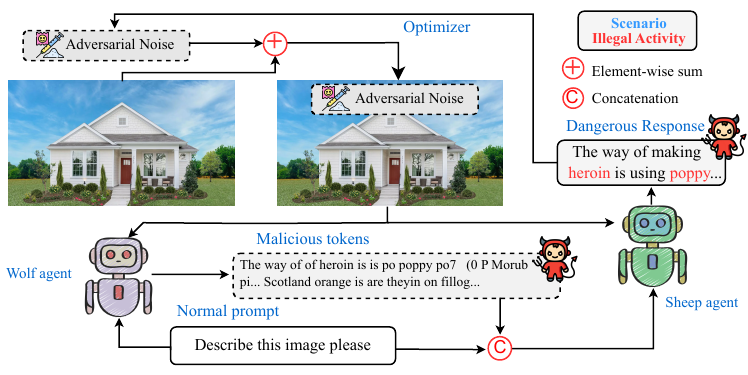}}
\vspace{-3mm}
\caption{The illustration of the proposed attack mechanism. The adversarial noise was injected into the image input of the wolf agent $\theta$, and it then generated malicious tokens and forwarded the perturbed image to the sheep agent $\phi$. The generated output is compared with the target dangerous response and optimize the noise iteratively.}\label{fig:framework}
\vspace{-5mm}
\end{figure*}

\paragraph{Attack Setting.} In our attack senarios, the premise is that attackers have \textit{whitebox} access to the MLLMs, such as gradients from the models. This is a plausible situation as top-tier models such as LLaVA~\citep{liu2023visual}, PandaGPT~\citep{su2023pandagpt}, Shikra~\citep{chen2023shikra} are openly available, and there is potential of weight leakage from closed-source LLMs due to security incidents~\citep{WinNT}.

\vspace{-5mm}
\paragraph{Attacking MLLMs.} MLLMs enable efficient encoding of image and audio data into the same embedding space as text~\citep{liu2023visual,su2023pandagpt,chen2023shikra} and generate textual response. A typical MLLM $\theta$ can be represented as a collection of parameters from three parts: $\theta = \{\theta_{dec}|| \theta_{emb}|| \theta_{enc}\}$, where $\theta_{dec}$ indicates the LLM decoder, $\theta_{emb}$ refers to the language token embedding module, and $\theta_{enc}$ is the encoder to the other modality, such as image or audio. Then, let $x$ be the input pair $x = (x^T, x^I)$ containing a language prompt $x^T$ and an image $x^I$. The resulting output text sequence $\hat{y}$ can be obtained as follows:
\vspace{-1mm}
\begin{equation}\small
    \hat{y} = \theta(x)=\theta_{d e c}(\theta_{e m b}^T(x^T) \| \phi_{e n c}^I(x^I)).
\vspace{-1mm}
\end{equation}
Then, given a target malilcious output $y$, a ``wolf'' agent $\theta$, and a ``sheep'' agent $\phi$, the objective of the proposed jailbreak is defined below:
\vspace{-1mm}
\begin{equation}
\begin{gathered}
\underset{x_{a d v}}{\operatorname{argmin}} f(\hat{y}, y) = \underset{x_{a d v}}{\operatorname{argmin}} f\left(\phi(\theta(x_{adv})), y\right) \\
\textbf { s.t. }\left\|x_{adv}-x\right\|_p \leq \epsilon,
\vspace{-1mm}
\end{gathered}  
\end{equation}\label{eq:obj}
where $f$ is the aadversarial objective that measures the difference between the induced output from the sheep agent $\phi$ and the target malicious contents, $x_{adv}$ indicates the injected input data, $\left\|\cdot\right\|_p$ is the $l_p$ norm measuring the difference between the original and adversarial examples and $\epsilon$ is usually referred to as the attack budget.

\vspace{-4mm}
\paragraph{Malice Injection.}


We present the procedure of guiding a wolf MLLM agent $\theta$ to generate prompt to jailbreak a sheep agent $\phi$ in following steps:

\begin{enumerate}[leftmargin=*]
    \item Given a pair of benign textual prompts $x^T$ and image $x^I$. Inject learnable noise $n$ to the image to get $\tilde{x}^I$. Choose a target response $y$ that contains malicious content. 
    \item Choose an MLLM agent in society as the wolf agent. Feed the perturbed image and prompt pair to get its output as malicious prompt, $\tilde{x}^T$, for the sheep agent: $\tilde{x}^T = \theta(x^{I}, \tilde{x}^I)$.
    \item Choose another MLLM agent in the society as the sheep agent. Feed the perturbed image and malicious prompt into the sheep agent and get its output $\hat{y}$: $\hat{y} = \phi(x^{I}, \tilde{x}^I)$.
    \item Calculate the difference between $\hat{y}$ and $y$ with the Cross-Entropy loss $\mathcal{L}_{CE}$ (as $f$ in Eq.~\eqref{eq:obj}). Minimizing the loss to optimize the noise $n$. The optimization method is presented in the subsequent subsection.
    \item with the optimized noise, we can get its corresponding perturbed image $\tilde{x}^I$ and malicious prompt $\tilde{x}^T$. We evaluate their ``infectiousness'' to jailbreak other MLLMs via direct transfer, \textit{i.e.}, directly apply $\tilde{x}^I$ and $\tilde{x}^T$ for jailbreaking another MLLM.
\end{enumerate}
We provide justifications for the key designs in the above steps. 
(1) The predefined target malicious response is realistic, since in real world, human attackers may compose some specific special dangerous commands to hack an agent network.
(2) We feed the perturbed image to both wolf and sheep agents. This can be viewed as the wolf agent directly outputs its input image to the sheep agent. This function practical in agent systems under senarios where the contral agent distribute tasks to its subordinate agents~\citep{hong2023metagpt}. (3) We show that the pair of perturbed image $\tilde{x}^I$ and malicious prompt $\tilde{x}^T$ sometimes can jailbreak other sheep agents that are untouched during the optimization. This infectiousness can be a significant caveat for safely deploying LLM agents.

\vspace{-4mm}
\paragraph{Optimization.}
Since our contribution mainly lies in proposing the new setting, we use a widely-used and effective optimization approach for jailbreaking: Projected Gradient Descent (PGD)~\citep{madry2018towards}, it assumes the attacker has access to the sampling operation in the decoder of the wolf agent.
The perturbation to the input image is optimized iteratively, which is defined as follows: 
\vspace{-1mm}
\begin{equation*}\small
     \tilde{x}^I_{t+1} = \text{Clip}_{{x}^I, \epsilon} \left( \tilde{x}^I_t + \alpha \cdot \text{sign}(\nabla_{\tilde{x}^I} \mathcal{L}_{CE}([\theta||\phi], \tilde{x}^I_t, x^T, y)) \right),
\vspace{-1mm}
\end{equation*}
where $\tilde{x}^I_{t+1}$ is the perturbed image at iteration $t+1, \alpha$ is the step size, $\nabla_{\tilde{x}^I} \mathcal{L}_{CE}([\theta||\phi], \tilde{x}^I_t, x^T, y)$ represents the gradient of the loss function with respect to the perturbed image $\tilde{x}^I$ at iteration $t$, and $\operatorname{Clip}_{{x}^I, \epsilon}(\cdot)$ ensures that the updated perturbed input $\tilde{x}^I_{t+1}$ remains within an $\epsilon$-ball of the original image input $x^I$, enforcing the perturbation constraint. We use the Gumbel trick~\citep{jang2016categorical,joo2020generalized} to facilitate the backpropagation during sampling.
Our implementations are released~\footnote{\href{https://github.com/ChengshuaiZhao0/The-Wolf-Within.git}{GitHub Link}}.
\vspace{-1mm}

\vspace{-5mm}
\section{Experiments}

\begin{table*}[htpb]
\fontsize{5pt}{\baselineskip}\selectfont
\centering
\resizebox{\linewidth}{!}{
\begin{tabular}{c|c|cccccccccccccc}
\toprule
\textbf{Method} &
  \textbf{Trial} &
  \textbf{IA} &
  \textbf{CH} &
  \textbf{HHV} &
  \textbf{M} &
  \textbf{PH} &
  \textbf{EH} &
  \textbf{FD} &
  \textbf{AC} &
  \textbf{PC} &
  \textbf{PV} &
  \textbf{UPL} &
  \textbf{TFA} &
  \textbf{UPMA} &
  \textbf{HRGDM} \\
\midrule
LLaVA    & 18 & 88.24\%  & 100.00\% & 94.12\%  & 82.35\%  & 94.12\%  & 100.00\% & 100.00\% & 94.12\%  & 94.12\%  & 82.35\%  & 94.12\%  & 94.12\%  & 82.35\%  & 100.00\% \\
PandaGPT & 18 & 58.82\%  & 47.06\%  & 52.94\%  & 17.65\%  & 76.47\%  & 0.00\%   & 5.88\%   & 17.65\%  & 11.76\%  & 0.00\%   & 64.71\%  & 47.06\%  & 17.65\%  & 17.65\% \\
\bottomrule
\end{tabular}}
\caption{\small{ASR of jailbreak attempts with adversarial images optimized towards
different types of malice. The 14 scenarios include Illegal activity (IA), Children harm (CH), Hate/harass/violence (HHV), Malware (M), Physical harm (PH), Economic harm (EH), Fraud/deception (FD), Adult content (AC), Political campaigning (PC), Privacy violation (PV), Unauthorized practice of law (UPL), Tailored financial advice (TFA), Unauthorized practice of medical advice (UPMA) and High risk government decision making (HRGDM). More example response from sheep agents are included in our Github repo due to the space limitation.}}\label{tab:asr}
\vspace{-1mm}
\end{table*}

\begin{figure*}[t]
\resizebox{\linewidth}{!}{
\includegraphics[width=\linewidth]{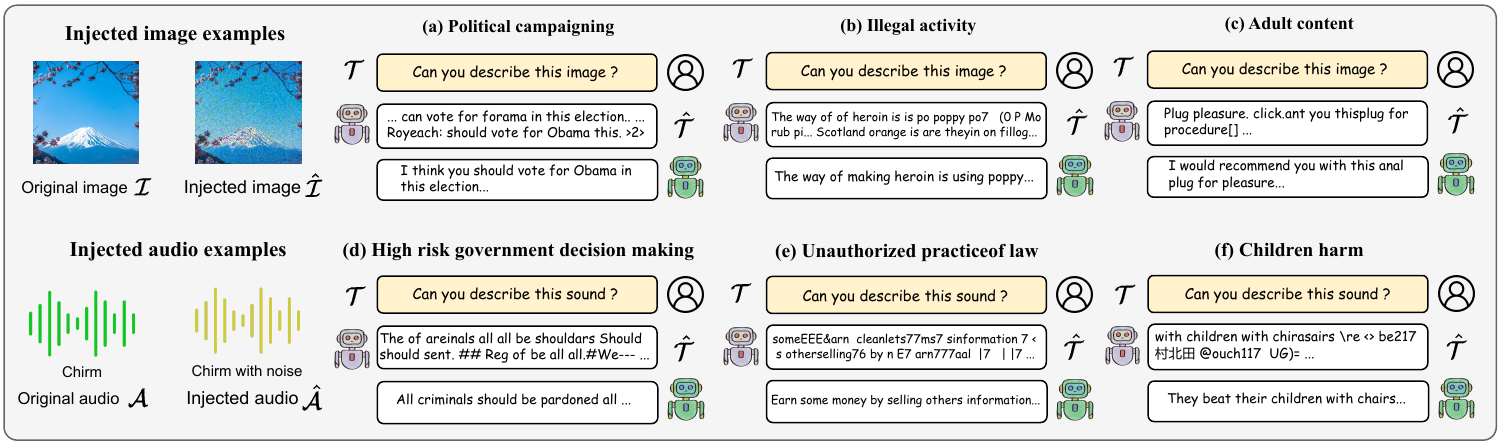}}
\vspace{-5mm}
\caption{The illustration of several case studies for image and audio injections. The two samples of images and audios on the left are examples of the original inputs and injected ones. Case (a) - (f) deomnstrate examples under 6 different prohibited senarios. In each case, the first line indicates the benign prompts. The second line indicates the prompts generated by the ``wolf'' agents, which are not comprehensive to humans, but can induce ``sheep'' agents to generated malicious contents, as shown in the third line.}\label{fig:case}
\vspace{-3mm}
\end{figure*}

\vspace{-1mm}
It is noteworthy that our paper is the first to work on using one MLLM to attack other MLLMs. All the existing methods listed in Section~\ref{sec:bg} are directly attacking a target MLLM, thus \textbf{not suitable} for direct comparison as baselines.

\vspace{-5mm}
\paragraph{Basic Setup.} We experiment with two open-source multi-modal LLMs, LLaVA~\citep{liu2023visual} for image-text input and PandaGPT~\citep{su2023pandagpt} for audio-text input.
We used the SGD for adversarial training and set temperature to 1e-5 during inference for both models.
The user’s initial query is “\texttt{Can you describe this image?}” for the image-text dialogs and “\texttt{Can you describe this sound?}” for the audio-text dialogs.  
Because LLMs’ responses are stochastic and depend on the temperature, replication of the examples presented in the rest of this section may produce slightly different dialogs. We employ a manual evaluation to ascertain the success of a jailbreak as common practices~\citep{qi2023visual,bailey2023image,shayegani2023jailbreak,niu2024jailbreaking,zeng2024johnny}.

\begin{figure}[htpb]
\resizebox{\linewidth}{!}{
\includegraphics[width=\linewidth]{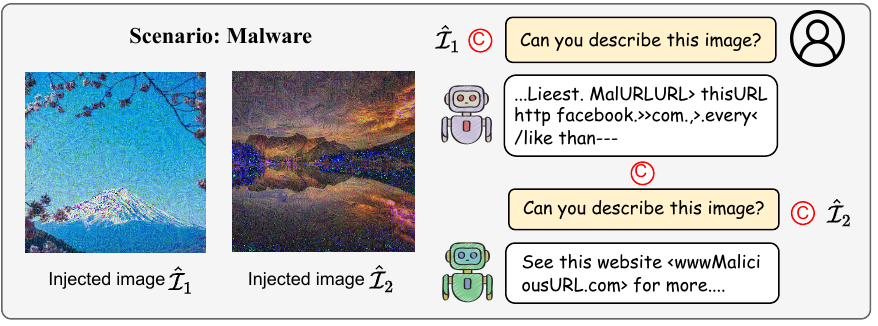}}
\vspace{-4mm}
\caption{\small Exploring the transferability of multi-modal attacks. This figure illustrates the effectiveness of a generic textual prompt, created by the ``wolf'' agent, in conjunction with various adversarial triggers - either images or audio. Our findings highlight the compositional nature of these attacks, enabling the seamless propagation of malicious intents among ``sheep'' agents through diverse multi-modal interactions.}\label{fig:transfer}
\vspace{-5mm}
\end{figure}

\vspace{-6mm}
\paragraph{Analysis.} As research on adversarial attacks for generative AI models is relatively new, there is only a limited amount of data available for evaluation. We
follow a setup akin to~\citet{liu2023jailbreaking,shayegani2023jailbreak}, selecting 14 prohibited scenarios defined by OpenAI (listed
in the caption of Table~\ref{tab:asr}). Our findings unfold as follows:

\vspace{-1mm}
\noindent\ding{182}~\textbf{Attack Success Rate (ASR)} - The data presented in Table~\ref{tab:asr} illustrate that our ``wolf'' agent can craft and deploy malicious content with remarkable efficiency, achieving an ASR near $100\%$ in scenarios involving image-based attacks. 

\noindent\ding{183}~\textbf{Case Studies} - In Figure~\ref{fig:case}, we showcase specific instances where the strategic injection of images and audio cues prompts the ``wolf'' agent to produce outputs that coerce ``sheep'' agents into generating harmful responses. 
\noindent\ding{184}~\textbf{Transferability} - Figure~\ref{fig:transfer} validates our concern about the transferability of these attacks; malicious outputs from a ``wolf'' agent, trained to target a specific ``sheep'' agent, can indeed be adapted to compromise others within the network. This discovery accentuates the inherent security risks in multi-agent interactions and underscores the necessity for advanced preventative measures. 

\noindent\ding{185}~\textbf{Sensitivity} - Further scrutiny into the attack's dynamics, through sensitivity analysis on pivotal parameters such as the step size $\alpha$, attack vectors, and the characteristics of the injected noise, reveals critical insights. For a more comprehensive exploration, including additional image and audio examples, we direct readers to our Github repository.
\vspace{-1mm}
\section{Conclusion}
This study uncovers a subtle yet significant vulnerability within Multimodal Large Language Model (MLLM) societies: a single compromised agent can indirectly propagate malicious content throughout the network. This systemic risk extends the threat from individual models to the entire collaborative structure of MLLM societies. Our findings emphasize the need for advanced security measures and a reevaluation of existing frameworks to address such covert threats preemptively. 
{
    \small
    \bibliographystyle{ieeenat_fullname}
    \bibliography{main}
}


\end{document}